# Pulsar skips: Understanding variations in the regular periods of rotating neutron stars.

Clayton Miller

Abstract: Pulsars are spinning neutron stars with very regular periods. These pulsars have, however, had instances where they exhibit a change in their periods. Older theories have shown that older pulsars have a tendency to skip and speed up. Newer theories have been created, due to the discovery that younger X-ray pulsars exhibit the same skips. The older theories explain that the core of the pulsar is a superfluid with a differential rotation and the core will occasionally exhibit solid properties to catch the crust of the pulsar and speed it up. The newer quantum mechanical theory states that quantum particle packets, called the strange nuggets, slam into the side of the pulsar to add angular momentum to the pulsar and then release it later.

I. Introduction: Neutron stars, intuitively named, are constructed of only neutrons and thus are very useful objects to observe in many physics fields. Since the discovery of Pulsars, they have been the focus of modern astrophysics. Pulsars are rotating neutron stars that emit a beam out of the poles that cause a flash in all the places where the beam hits. If that beam hits earth we can see this flash and it is what makes pulsars look like lighthouses, constantly spinning and flashing light in our direction. This "lighthouse effect" is confined to very regular intervals. Each pulsar shines its light toward Earth with unique timing; occasionally, its intervals will speed up briefly before resettling back into its original, regular schedule. The idea of speeding up has not been advanced much in previous years because it is incredibly difficult to observe these celestial bodies since we can only observe the ones that emit light towards the earth. The leading theory is that pulsars have a superfluid core that has a differential rotation that is corotating with the crust that prevailed. The superfluid theory continues to exhibit gaps in knowledge and there was never enough proof to get a decent idea if it was true.

In 1999, We were able to see higher energy pulsars exhibiting the same skipping anomalies from new technology, such as the Chandrasekhar and Hubble space telescopes.[10] The superfluid model suggests that the skipping only happened in the Vela radio pulsar, and the presence of these high energy pulsars skips has led us closer to understanding the skipping anomaly. With improved telescopes being commissioned soon, we will be able to better observe pulsars. Telescopes like Laser Interferometer Gravitational-Wave Observatory(LIGO), Imaging X-ray Polarimetry Explorer( IXPE), Lunar Ultraviolet Cosmic Imager(LUCI), and the long-delayed James Webb telescope (JWST), will detect pulsars of different energy levels. With different energy levels in pulsars, we will see if this theory holds and also get better data surrounding the skips. This paper will review the theories of neutron stars that may lead them to experience these "skips" and what other observation techniques will be applicable.

II. Pulsar Formation and Classification: Pulsars are one of the most condensed objects in the universe that are created by rotating neutron stars. The condensed matter that builds celestial bodies, like neutron stars, tends to have some interesting make-up and features, specifically noting that they violate specific laws. After black holes, neutron stars are the second most condensed objects in the universe. They exhibit properties, unlike any other celestial body, making them important to modern astronomy and physics. Neutron stars are formed in supernovae, the most energetic explosions known. After the star explodes, the remnant of the core is a neutron star. The angular momentum and magnetic flux of the star are both conserved within the neutron star.[10] This leads to the increase of rotational speed in the neutron star until it speeds up enough and creates a pulsar[10]. Neutron stars in a more up-close picture show some very interesting characteristics, like their violation of the Pauli- exclusion principle (PEP). The PEP is defined as two fermions that cannot occupy the same state at the same time[1]. The violation of this natural principle means that the matter that makes up neutron stars is comprised only of neutrons. The protons and electrons that once helped build the predecessor of the neutron star fuse and release energy but the remnants are just neutrons. Misleadingly the name implies that the object is indeed a star, which is not true because all the energy it emits is residual and is no longer performing nuclear fusion. All this energy can be emitted into beams due to the neutron star's immense magnetic field. If the beams of energy or magnetic poles are misaligned with the axis of rotation, much like Earth's, then we can experience a sort of " lighthouse" effect, where every time the beam passes by the observer we see a flash of light. These flashes are set at very regular intervals like once per second or a thousand times per second. These intervals are extremely regular, but will very rarely break this regularity, as will be explained in the rest of this paper.

      Pulsars are mostly classified by the catalog of LGM, BeXB, and others. These categories of pulsar have allowed us to study different specific aspects of the regular pulsar by exposing its extremes. The magnetar, for example, demonstrates magnetic fields, that are unparalleled in the universe, hence its name. Regardless of the type of pulsar, they all are recognizable by their

distinct and set rotational periods, but these periods sometimes have mistakes. This mistake is shown by the pulsar's rotational speed increasing for a period of weeks and then settling back into the original speed.

Leading theories describing the changes in speed explain that the core of neutron stars is made up of a superfluid substance that has a rotational rate that differs from the crust of the neutron star, and quantum particles slam into the side of the neutron star speeding it up and causing it to "skip".

III.Properties: The properties of neutron stars, superfluids, and strange nuggets are very important to understanding the pulsar skips. Properties of these include their makeup and specific characteristics that allow them to exert these phenomena.

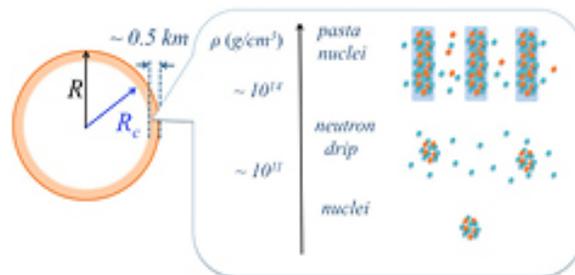
.

Diagram of the density of a neutron star, by describing the density of individual particles. (Gordon Baym *et al* 2018 *Rep. Prog. Phys.* **81** 056902)

III.1 Superfluid: Superfluids are classified as a non-newtonian matter which means that they do not abide by classical Newtonian rules. Friction falls into this category that superfluids do not follow so does their matter state. Superfluid traits are what lead many to believe that is the key to understanding the reason for pulsar skips. Oobleck is one of the best-known superfluids. Many people have experienced these interesting characteristics, like when it is squeezed it becomes a solid, but immediately before that happens it is liquid. One of the theories described in this paper will indicate that it is possible for the core of a neutron star and therefore the core of a pulsar to be a superfluid. However, this superfluid may have characteristics beyond what we know because it is in a condensed matter object with properties we are still not familiar with. Within these stars, we have to expect that the superfluid core will act with relativistic properties because of the quarks' non-significant masses. These relativistic properties could also be a reason that neutron stars skip.

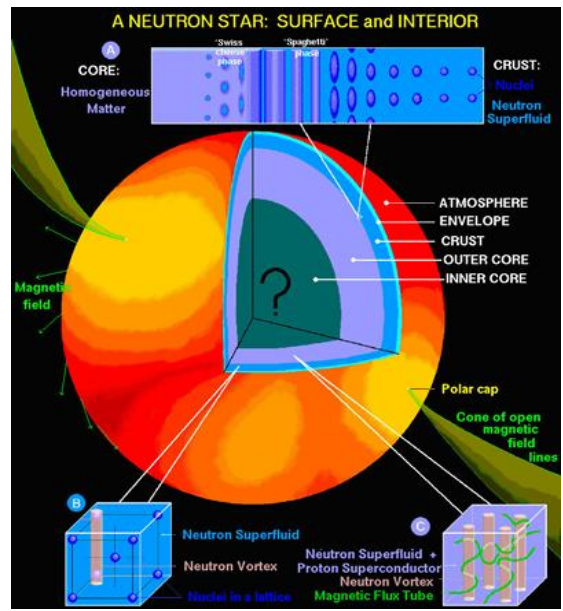

(Depiction of the neutron star with the theorized superfluid substance at the core
https://www.southampton.ac.uk/maths/research/projects/superfluidity.page)

III.2 Strange Quarks and Nuggets: Strange nuggets are large collections of strange quarks. Strange quarks are quantum particles with a charge of -⅓ and spin of ½. Strange Nuggets or strangelets are theorized and have not been observed. These large packets of strange quarks may be non-baryonic matter or dark matter, which could be why they have never been detected.

Even though the matter is only a theory it is still one of the main explanations of the pulsars skips.

      III.3 Formation of Neutron Stars: Neutron stars are formed in supernovae, the most energetic explosions in the universe. Supernovae have the energy of $10^{29}$ of the most energetic bombs ever created by man. The remnants of the supernova are neutron stars. Neutron stars are essentially the corpses of massive stars that have a mass of 1.4 solar masses. Slightly less deadly cousin of the black hole.

      III.4 Formation of Pulsars: Pulsars are formed as the neutron stars form. Neutron stars start out as stars larger than our own sun and after a supernova, they end up the size of Boulder, Colorado. We can safely conclude that angular momentum is conserved since no outside toques have acted on the star before or after the supernova. Using the basic definition of angular momentum. $L = r_s \times p$ where $p = mv$, $r_s$ is the radius of the star, and angular momentum is *L*, *m* is the mass of the neutron star, and *v* is the velocity tangential to the surface of the neutron star. After the supernova, a large amount of mass, m, is held in the neutron star. So, m is still very large, but r has decreased significantly to be r thus the value of v must increase significantly to keep the same *L*. Therefore, *v* is inversely proportional to *r*. As the radius decreases the velocity increases, much like the classic figure skater example. As the figure skater pulls in their arms while spinning, they start to spin faster, the same concept is what creates a pulsar.

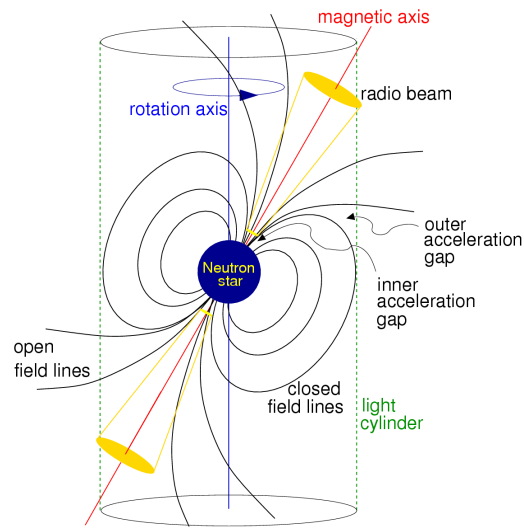

Diagram of a pulsar's typical motion and orientation (https://www.cv.nrao.edu/course/astr534/Pulsars.html)

V. Glitch Behavior: The glitch behavior in pulsars remains one of the greatest mysteries of the already mysterious bodies. As shown in the diagram below, pulsars have a period that is faster than the original one measured by astronomers. The Vela pulsar is one of the more studied pulsars because of its easy measurement. The Vela pulsar's glitches were discovered in 2006 and theorized to be the result of gravitational waves. These gravitational waves are described as lasting "hours to months after the two glitches from the Vela pulsar", but were removed as an explanation after no significant measurements were made by LIGO.[13]

VI. Theories: The theories to explain these pulsar skips consist of explaining using classical and quantum mechanical approaches. Both explain the phenomena with similar detail, however, both have their respective drawbacks.

VI.1 Superfluidity/ Angular momentum transfer: The theory that explains skips using the superfluidity of the core does so by explaining that the core is superfluid and also has a rotational speed that is greater than the crust. Since the superfluid core rotates faster than the crust when it exhibits solid properties it will essentially grab the crust and speed it up to the speed that the core is at. This is boiled down to basic angular momentum transfer and is the result of not only the difference in angular velocity but also inertia. Described in the equations below.[3]

$$I_{total} = \frac{8\pi}{3} \int_0^R dr r^4 \frac{(\varepsilon(r) + P(r))}{\sqrt{1 - 2GM(r)/r}} \frac{\bar{\omega}(r)}{\Omega} e^{-\nu(r)}.$$

$$I_{crust} = \frac{8\pi}{3} \int_{R_c}^R dr r^4 \frac{(\varepsilon(r) + P(r))}{\sqrt{1 - 2GM(r)/r}} \frac{\bar{\omega}(r)}{\Omega} e^{-\nu(r)}$$

The ratio of equation (1) and (2) becomes.

$$\frac{I_{crust}}{I_{total}} = \frac{I_{crsf} + I_{crnsf}}{I_{crsf} + I_{rest}}$$
$$\frac{I_{crust}}{I_{total}} > \frac{1}{1 + I_{rest}/I_{crsf}}$$

This shows that the inertia of the two parts of the pulsar are different and therefore the angular momentum will differ continuing the theory of superfluid angular momentum transfer.

The superfluid theory applies only to the older generations of neutron stars.[2] The theory states that since the neutron star loses rotational energy over time due to emission the crust will start to slow, while the core remains near the original rotational speed. The superfluidity of the core allows it to sometimes "catch" the core to create the spin-up of the pulsars. This theory has since been disproven, at least for a few reasons, due to observations of x-ray pulsars. The x-ray pulsars are younger pulsars that emit higher energies of light and therefore a shorter wavelength.[11] This new energy can be explained by the massive amounts of energy that the pulsar still has from the explosion of its mother star. These x-ray pulsars have also been observed with glitches. Since the theory about glitches was oriented towards older pulsars, it must be modified to accommodate these new observations. Since the theory was wrong in one aspect, of course, this leaves the rest of it open to skepticism and critique. A newer theory calls for "strange nuggets", which is a fundamental particle that would interact with the surface of a pulsar and cause it to expel energy in its elastic form. (4). This quantum particle slams into the side of the pulsars crust causing a sudden jump up in the rotational speed. (4) This theory proves to be more resilient in the face of the age of pulsars. The only fall back of it is those strange nuggets, themselves, and their existence is still somewhat up for debate. These two theories have been competing in this niche field for a number of years with no real progress in either field making it hard to determine a better choice.

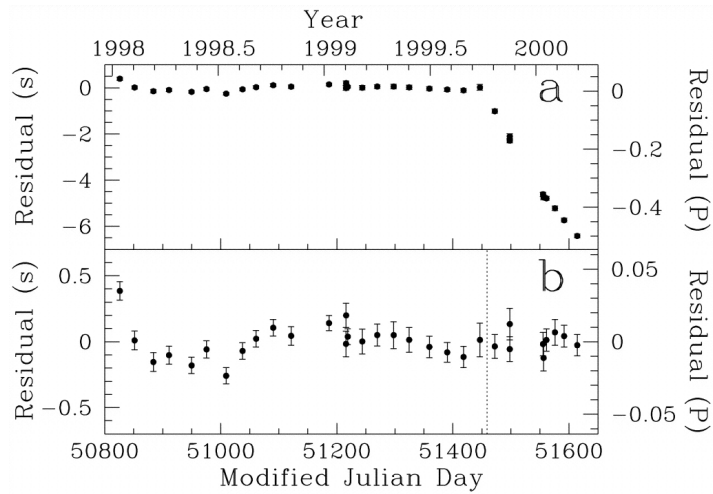

Dataset depicting the pulsar regular rotational rate being disrupted somehow. (A Glitch in an Anomalous X-Ray Pulsar (Victoria M. Kaspi,[11] Jessica R. Lackey,[11] and Deepto Chakrabarty [11.]))

VI.2 Stange Nuggets Theory: Strange Nuggets are quantum packets, which are basically clusters of strange quarks, that create skips by slamming into the sides of pulsars along with the rotation to increase the speed temporarily after then the pulsar releases this energy and then returns to its normal frequency. The greatest issue of this theory is that the strange quarks won't always collide in the same direction of the rotation of the neutron star. The neutron star receives the energy from the strange quark and then releases it. Strange nuggets must have the direction of the velocity as the linear component of the rotational energy. This theory allows for the decrease of rotational speed but also has its own setbacks. This setback is that the quantum mechanical quarks do not have an exact direction at all times. Since they do not necessarily move in the same direction that means that the strange nugget theory is missing a very key detail, the direction of the strange nuggets may not be the same as the vector tangent to the rotational velocity.

VII. Methods of Observation: Methods of observation regarding pulsars have been mostly confined to radio astronomy because it was how they were first discovered and have not changed much since then. Radio wavelengths are visible on the earth's surface, but it tends to be weaker in energy. X-ray is a newer observation method for pulsar observation but it must be observed from space and is much more energetic to observe and therefore more visible.

VII.1 Using Radio: Pulsars have been observed in the radio spectrum for many years, however, there is not a particular reason for this. Jocelyn Bell Burnell was the first to notice these flashes while working as a data analyst in 1967. Mistakenly, she assumed that these regular radio signals were sent by aliens, which she quickly named little green men (LGM). LGM remained the dominant cataloging database for pulsars for a number of years, like many things in astronomy, due to tradition. LGM slowly changed into many other catalogs depending on recently discovered subcategories of pulsars, like magnetars or soft gamma pulsars.

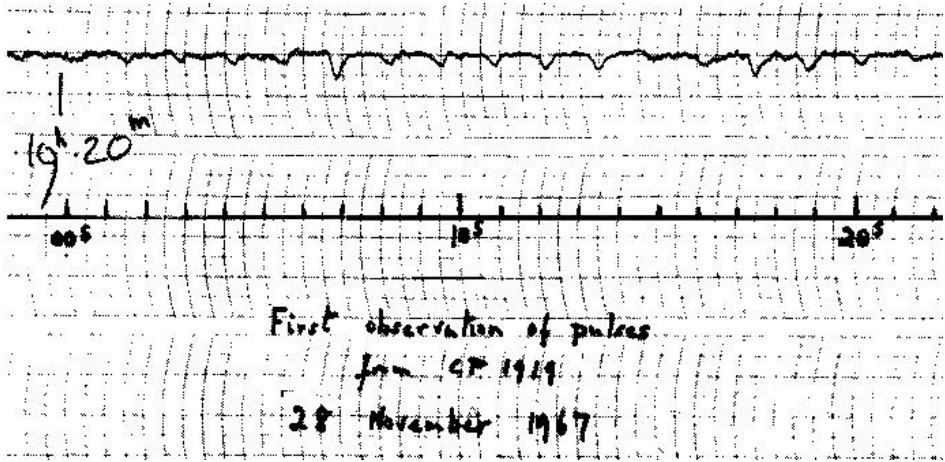
(https://www.cv.nrao.edu/course/astr534/Pulsars.html) Jocelyn Bell Burnell's initial observation leading to the discovery of pulsars.

VII.2 Using X-ray as outliers: X-rays have become a great way to observe and measure pulsars because has a short wavelength on the electromagnetic spectrum and are therefore more energetic. Since the discovery of skips in X-rays it has become significantly more viable than it was in the past to discover the skips. Chandrasekar is the current pinnacle of X-ray space telescope technology, but it has not been advanced enough to make any significant progress in the study of pulsar skips.

VIII. Conclusion: While these theories may be good ideas, it has become reliant on technology to catch up with the requirement of the theories. The more the technology is developed, the more will be discovered about these pulsar skips. Space telescopes are the primary method that will allow the concepts of pulsar skips using superfluid angular momentum transfer or using strange nuggets to increase the angular momentum by hitting the pulsar along the tangential velocity vector.

VIII.1 Advancement: Better space telescopes are on the way and will advance the study of pulsar skips substantially. ILO-1 is the international lunar observatory that will be placed on the moons and observe the pulsars in several wavelengths that will all benefit from the lack of atmosphere, light, and stability of the moon. XRISM is a space-based X-ray telescope that will help detect and observe the younger pulsars to obtain more evidence that skips occurring in younger pulsars.

VIII.2 Discussion: Pulsars are not just inherently interesting, they serve a purpose. Studying pulsars allows the astronomic community to learn more about condensed matter objects occurring in nature. A condensed matter study allows researchers to better understand the extent of general relativity. Learning about the natural limits of general relativity will give us the foresight into cautions and methods to possibly create our own condensed matter. Understanding these unique objects will help humanity work to better understand the gravity and its effects in stronger magnitudes.